\documentclass[10pt,conference]{IEEEtran}
\usepackage{cite}
\usepackage{booktabs}
\usepackage{graphicx}
\usepackage{amsmath}
\usepackage{array}
\usepackage{microtype}
\usepackage{xcolor}
\usepackage[hidelinks]{hyperref}
\usepackage{tikz}
\usetikzlibrary{positioning,arrows.meta,calc}
\newcommand{\SUCCESS}{\ensuremath{\text{\textsc{success}}}}
\newcommand{\FAILINC}{\ensuremath{\text{\textsc{failure\_or\_incomplete}}}}
\newcommand{\UNCLEAR}{\ensuremath{\text{\textsc{unclear}}}}
\newcommand{\COMPLETED}{\ensuremath{\text{\textsc{completed}}}}
\newcommand{\NOTCOMPLETED}{\ensuremath{\text{\textsc{not\_completed}}}}
\newcommand{\UNKNOWN}{\ensuremath{\text{\textsc{unknown}}}}
\newcommand{\UNSURE}{\ensuremath{\text{\textsc{unsure}}}}
\newcommand{\PASS}{\ensuremath{\text{\textsc{pass}}}}
\newcommand{\FLAG}{\ensuremath{\text{\textsc{flag}}}}
\newcommand{\REVIEW}{\ensuremath{\text{\textsc{review}}}}
\newcommand{\YES}{\ensuremath{\text{\textsc{yes}}}}
\newcommand{\NO}{\ensuremath{\text{\textsc{no}}}}
\newcommand{\NA}{\ensuremath{\text{\textsc{not\_applicable}}}}

\begin{document}
\title{Who Verifies the Agent's Success Claim? Self-Report Leakage in GUI-Agent Completion Judging}
\author{
\IEEEauthorblockN{
Yunfeng Zhu$^1$,
Shengcheng Yu$^{2,*}$,
Zihan Li$^3$,
and Zhenyu Chen$^1$
}
\IEEEauthorblockA{
\textit{$^1$State Key Laboratory for Novel Software Technology, Nanjing University}, Nanjing, China \\
\textit{$^2$Technical University of Munich}, Heilbronn, Germany \\
\textit{$^3$Nanjing Normal University}, Nanjing, China \\
yunfengzhu@smail.nju.edu.cn, shengcheng.yu@tum.de, \\
lizihan@njnu.edu.cn, zychen@nju.edu.cn \\
$^*$Corresponding author: Shengcheng Yu
}
}

\maketitle

\begin{abstract}
In automation pipelines, a GUI agent's confident success report can let downstream stages proceed even when the recorded execution trace does not show that the agent produced the requested outcome. We call this case a \emph{dangerous misreport}: an unhedged success claim that the trace does not support as causal accomplishment during the run. We study \emph{report--trace auditing}, which checks an agent's completion claim against an independently recorded trace.

A controlled \textsc{aware}/\textsc{blind} intervention reveals \emph{self-report leakage}: the agent's own self-report changes how completion is judged. The same judge reads the same task and trace; only that self-report is included or withheld. Withholding it improves trace-only dangerous-misreport recall by 43.4 percentage points with Qwen3.5-35B-A3B and 13.3 percentage points with DeepSeek-284B-A13B, with the same trend on held-out Batch~2. This motivates \emph{report isolation}. The report records what the agent claims, but it must not influence the trace-state judgment. 

We implement this separation with a structured auditor that separates claim interpretation from claim-blind trace analysis and applies deterministic rules to extracted evidence. Because evidence extraction is separated from the policy that consumes it, the same audit record can be replayed under alternative policies without new model calls.

The auditor achieves F1 0.75--0.78 on the 465-record audit-target set and 0.77--0.81 on held-out Batch~2, outperforming claim-aware and \textsc{prompt-blind} controls in recall and F1. Under audit-stage gating, where \textsc{flag} and \textsc{review} both block automatic pass, it blocks 74--90\% of dangerous misreports at 45--56\% inspection load, or 16--19\% of the full in-scope stream after coarse status triage.
\end{abstract}


\section{Introduction}
\label{sec:intro}

LLM-driven GUI agents are increasingly used inside mobile testing and regression workflows~\cite{liu2024gptdroid,hu2024auitestagent,yoon2024intent,kong2025prophetagent}, where task outcomes may be consumed by downstream harnesses or review pipelines. Consider a run that ends with \emph{``Done---the duplicates were removed.''} The harness records the task as \textsc{pass}, downstream stages proceed, and the resulting device state enters the pipeline as if the requested operation had succeeded. But the trace may show a different execution: the agent inspected an empty list, acted on the wrong screen, or failed to save the requested change. No exception is thrown. The failure remains silent because the agent under audit supplies its own success report.

We study report--trace auditing. Given a completed run, the agent reports completion and the audit checks whether the independently recorded trace supports that report. A \emph{dangerous misreport} occurs when the report claims success but the trace does not show causal accomplishment, meaning that it does not show the agent itself producing the task-relevant outcome during the run. Final-screen compatibility alone is insufficient, because pre-existing goal states, empty target lists, proxy screens, and unsaved edits can all make a failed run appear successful~\cite{ye2025claweval,rui2026finish}. Such cases are dangerous because they can be recorded as silent false pass outcomes.

Claim-aware completion judging creates a circular evaluation path in GUI-agent evaluation. Because task-specific success checks are costly to author across heterogeneous GUI tasks and human trajectory review is costly, recent agent-evaluation work increasingly uses LLM or agent judges for trajectory success~\cite{barr2015oracle,molina2025oracle,zhuge2025agentjudge,lu2025agentrewardbench}. A common pattern is to ask the judge to read the task, the execution trace, and the agent's final report in a single prompt, then decide whether the task was completed~\cite{pan2024autonomous,ma2024agentboard,xue2025illusion,lu2025agentrewardbench,he2024webvoyager}. The report thereby ceases to be merely a claim to be verified and instead becomes part of the evidence for its own truth, making the evaluation path circular.

We quantify this circularity with an \textsc{aware}/\textsc{blind} intervention. The same completion judge reads the same task and trace; only the agent's self-report is included or withheld. If the self-report were harmless context, the trace-state judgment would remain stable. It does not. Withholding it raises trace-only dangerous-misreport recall by 43.4 percentage points with Qwen3.5-35B-A3B and 13.3 percentage points with DeepSeek-284B-A13B on the combined audit-target set, and the held-out Batch~2 split shows the same trend. Exposing the self-report shifts trace judging toward completion and suppresses dangerous-misreport recall. These results make \emph{report isolation} a design requirement: the report records what the agent claims, but it must not shape the judgment of the trace.

Report isolation alone is not sufficient. We therefore build a report-isolated evidence--policy auditor. A \textsc{prompt-blind} judge withholds the report, but it still returns an opaque verdict that is difficult to inspect or reuse under a different completion policy. Our auditor instead separates the problem into evidence extraction and policy enforcement. One component reads the report to identify the claim; a second reads only the task and trace to extract typed evidence of accomplishment. Neither decides completion directly. Deterministic rules consume the extracted evidence and assign each run to \textsc{pass}, \textsc{flag}, or \textsc{review}. This evidence--policy boundary also makes the audit replayable: cached evidence can be evaluated under alternative policies without another model call~\cite{huang2010leap,huang2013clap}. The output is not a prompt-level verdict but a software audit record whose evidence fields and policy decisions can be inspected, regression-tested, and replayed.

We evaluate this auditor on 465 audit-stage records drawn from 1{,}384 in-scope GUI-agent execution attempts, including 180 dangerous misreports. On the full audit-target set, the auditor achieves F1 0.75--0.78, and on held-out Batch~2 it reaches 0.77--0.81, outperforming claim-aware and \textsc{prompt-blind} controls in recall and F1. Under audit-stage gating, where both \textsc{flag} and \textsc{review} block automatic pass, it blocks 74--90\% of dangerous misreports from automatic pass at 45--56\% inspection load within this audit-target set.

This paper makes three contributions:
\begin{enumerate}
\item \textbf{We quantify self-report leakage in GUI-agent completion judging.}
Holding the task, trace, model, serialization, and scoring rule fixed, exposing the agent's self-report shifts trace-state judgments toward completion and suppresses dangerous-misreport recall. The result makes report isolation an empirical design constraint rather than an optional prompt choice.

\item \textbf{We define dangerous misreporting by separating the report from trace evidence.}
The formulation separates what the agent claims from what the trace supports. Claim intent is labeled from the report alone, while causal accomplishment is labeled from the task and trace alone. Thus, the claim that triggers the audit target is not used to judge trace completion.

\item \textbf{We enforce report isolation through a replayable evidence--policy boundary.}
The auditor separates claim reading, claim-blind evidence extraction, and deterministic policy enforcement, producing an inspectable audit record that can be replayed under alternative policies without new model calls. This turns report isolation from a prompt-level instruction into a software-enforced information-flow and policy boundary, while improving dangerous-misreport recall and F1 over both claim-aware and \textsc{prompt-blind} controls.
\end{enumerate}
\section{Problem Formulation}
\label{sec:formulation}

This section formalizes report--trace auditing for a completed GUI-agent run. The formulation separates three questions that one-shot claim-aware judging often conflates: what the agent claims, what the trace supports, and whether the claim is consistent with the trace.

\subsection{Audit-Target Runs and Completion Construct}

We study post-hoc auditing of a GUI-agent run
\[
u=(x,\tau,m),
\]
where \(x\) is the natural-language task, \(\tau\) is the recorded execution trace, and \(m\) is the agent's final report. In this section, \(m\) denotes only the final report. More broadly, we use \emph{self-report} for the agent's own assertions about its progress, which occur at two granularities: the final report \(m\), a single end-of-run completion assertion, and \emph{per-step self-reports}, one short completion or progress claim the agent emits after each recorded step (e.g., ``deleted recipe~X''). Per-step self-reports are agent-authored claims---not chain-of-thought reasoning, and not part of the recorded GUI states in \(\tau\)---stating what the agent asserts it did, not why it acted. The final report \(m\) is the sole source of the claim-intent label \(c^*\) introduced below; per-step self-reports never assign \(c^*\) and enter the analysis only as additional agent-authored context in the \textsc{aware} condition (Section~\ref{sec:rq1}).

The final report is the claim source, while the trace is the independent evidence source. Since the report is produced by the same agent whose execution is under audit, it conveys what the agent claims. The trace, by contrast, supplies independent evidence of whether the claimed accomplishment occurred.

We model the trace as
\[
\tau=\langle s_0,s_1,\ldots,s_n\rangle,
\]
where \(s_i\) denotes the observable GUI state after the \(i\)-th recorded step. Each state contains GUI information such as visible text, content descriptions, resource identifiers, and bounds. The auditor operates offline from a completed trace, without access to a live device or an executable task-specific oracle, unlike benchmarks that provide environment checkers~\cite{rawles2025androidworld,jimenez2024swebench,xie2024osworld,xu2025androidlab}.

A GUI task can be judged complete under different semantics, including final-state satisfaction, causal accomplishment, and report faithfulness. We use \emph{causal accomplishment} as an audit construct: the trace must show evidence that the agent performed a task-relevant action and reached the requested post-condition during the run, rather than merely ending in a compatible state. A goal state already present at \(s_0\), an empty target list, or a typed but unsaved value is therefore not by itself evidence that the agent accomplished the task.

Let
\[
y^*(x,\tau) \in
\{ \COMPLETED,\NOTCOMPLETED,\UNSURE \}
\]
denote the human gold completion judgment under this causal-accomplishment policy. It is assigned based on the task and trace without revealing the final report to annotators.

\subsection{Claim Intent, Trace Support, and Positive Class}

A report--trace auditor separates claim interpretation from trace-state estimation. The claim-intent function
\[
C(m) \in
\{ \SUCCESS,\FAILINC,\UNCLEAR \}
\]
captures what the agent says, not whether the claim is true; \FAILINC{} denotes failure or incompletion. The trace-state function
\[
T(x,\tau) \in
\{ \COMPLETED,\NOTCOMPLETED,\UNKNOWN \}
\]
returns what the trace supports under the audit policy. A valid auditor must keep this trace-state estimate claim-blind: the input to the trace-state estimator excludes \(m\), so changing the final report may change \(C(m)\), but cannot change \(T(x,\tau)\). This information-flow requirement is the key difference between report--trace auditing and one-shot claim-aware judging.

The ground truth is defined over two independently annotated axes. The claim label
\[
c^* \in
\{ \SUCCESS,\FAILINC,\UNCLEAR \}
\]
is derived from the final report alone, while the actual-completion label \(y^*\) is derived from the task and trace alone, under the claim-blind causal-accomplishment policy. A record is a dangerous misreport iff
\[
c^*=\SUCCESS
\quad\wedge\quad
y^*=\NOTCOMPLETED.
\]
This is the positive class for evaluation: a failed execution risks being misclassified as \PASS{} because the agent reports success. Self-reported failures, where the report also indicates failure or incompletion, are not dangerous misreports because they do not create a silent false pass. The reverse inconsistency, where the report indicates failure or incompletion but the trace supports completion, may matter operationally as an underclaim, but it is not the positive class studied here.

The implemented detector predicts a dangerous misreport iff
\[
\begin{aligned}
\widehat{d}(x,\tau,m)=1
\;\Longleftrightarrow\;
& C(m)=\SUCCESS \\
& \wedge\; T(x,\tau)=\NOTCOMPLETED.
\end{aligned}
\]
Labels and predictions are not circularly dependent: gold positives use human labels \(c^*\) and \(y^*\), while predictions use the implemented pipeline. Because prediction depends on \(C(m)\), Section~\ref{sec:evaluation} also reports a gold-claim condition that replaces \(C(m)\) with \(c^*\) to isolate trace-channel behavior.

\subsection{Audit Disposition and Evaluation Views}

The auditor returns a disposition
\[
A(C,T) \in \{\PASS,\FLAG,\REVIEW\}.
\]
The overclaim case corresponds to a dangerous misreport:
\[
C(m)=\SUCCESS
\;\wedge\;
T(x,\tau)=\NOTCOMPLETED
\;\Rightarrow\;
\FLAG.
\]
Uncertainty in either channel routes the record to \REVIEW; agreement between the claim channel and checked trace state routes it to \PASS. Here \PASS{} is an audit disposition: it means that the report--trace consistency check does not flag the run as a dangerous misreport, not that the underlying GUI task is necessarily completed.

We report two evaluation views. In the automatic-detection view, only \FLAG{} counts as a positive prediction and \REVIEW{} counts as a non-detection. In the deployment-gating view---\emph{audit-stage gating}---both \FLAG{} and \REVIEW{} block automatic pass, because uncertain runs should not be silently accepted by an automation pipeline.

\section{Report-Isolated Evidence--Policy Auditor}
\label{sec:design}

Section~\ref{sec:formulation} defined report--trace auditing as a comparison between what the agent claims and what the execution trace supports. We implement this formulation as a structured report-isolated auditor:
\[
D(x,\tau,m)=A\bigl(C(m),G(x,E(x,\tau))\bigr),
\]
where \(C\) reads the final report, \(E\) extracts claim-blind evidence from the task and trace, \(G\) maps that evidence to a checked trace state, and \(A\) maps the claim--trace pair to an audit disposition. The rest of this section defines each component in this composition.

\begin{figure}[!t]
\centering
\includegraphics[width=0.9\columnwidth]{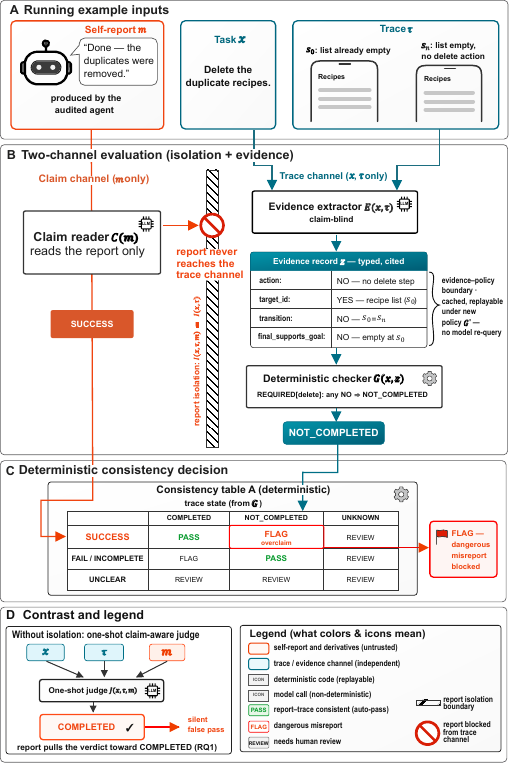}
\caption{Report-isolated evidence--policy auditing on a running
vacuous-delete example. (A)~Inputs: task \(x\), trace \(\tau\), and the
audited agent's final report \(m\). (B)~The report crosses the dashed
isolation boundary only through the claim reader \(C\); the claim-blind
extractor \(E\) receives only the task and trace, emits typed and cited
evidence \(z\), and the deterministic checker \(G\) maps \(z\) to a
trace state. (C)~The deterministic table \(A\) maps the overclaim case
(\SUCCESS{} claim, \NOTCOMPLETED{} trace state) to \FLAG. (D)~Without
isolation, a one-shot claim-aware judge can consume the same report and
silently pass the run; cached \(z\) replays under alternative policies
\(G'\) with no new model calls.}
\label{fig:structured-arch}
\end{figure}

The architecture is driven by two goals: enforcing the information-flow requirement in Section~\ref{sec:formulation}, and exposing an inspectable, replayable evidence--policy boundary that a prompt-level verdict cannot provide. The central design choice is not to ask a model for a final completion verdict. Instead, the model emits structured evidence fields from the task and trace, and deterministic code decides whether those fields satisfy the audit policy. The auditor extracts evidence once, checks it against deterministic rules, and caches the result so that changing the policy does not require new model calls. Figure~\ref{fig:structured-arch} shows the architecture for one audit-target run \(u=(x,\tau,m)\).

\subsection{Structured Evidence Interface}

The claim reader \(C\) reads the final report and classifies it as \SUCCESS, \FAILINC, or \UNCLEAR. A report is classified as \SUCCESS{} only when it conveys an unhedged assertion that the requested task was completed. Reports that state failure, partial completion, inability, uncertainty, or ambiguous status are not treated as confident success. The claim reader is a single temperature-0 model call per record, parsed into this three-way label space; the same claim reading is shared across backends and baselines, so paired comparisons isolate how trace support is estimated.

The trace channel begins with an evidence extractor:
\[
E(x,\tau)=z.
\]
The extractor reads the task and execution trace, but not the agent's self-report. It emits a structured evidence record:
\[
z=
\{e_{\mathrm{act}},e_{\mathrm{id}},e_{\mathrm{trans}},
e_{\mathrm{commit}},e_{\mathrm{final}}\}.
\]
The five evidence fields correspond to the obligations needed to audit causal accomplishment: whether a task-relevant action was observed (\texttt{action}), whether the correct target was involved (\texttt{target\_id}), whether a goal-directed state transition occurred (\texttt{transition}), whether the relevant change was committed or made durable (\texttt{committed}), and whether the final or post-action state directly supports the task post-condition (\texttt{final\_supports\_goal}). These identifiers are the field names used by the deterministic checker below.

Each evidence field is represented as
\[
\begin{aligned}
&e_i=(v_i,p_i,h_i),\\
&v_i\in
\{\YES,\NO,\UNKNOWN,\NA\},
\end{aligned}
\]
where \(v_i\) is the field value, \(p_i\) is cited trace support such as a step, screen, or span, and \(h_i\) is a short rationale. The extractor is not allowed to output \COMPLETED{} or \NOTCOMPLETED{}; its output is an auditable intermediate representation consumed by deterministic code.

\subsection{Deterministic Trace-State Checking}

The deterministic checker maps the task and evidence record to a checked trace state:
\[
G(x,z) \in
\{\COMPLETED,\NOTCOMPLETED,\UNKNOWN\}.
\]
It implements the causal-accomplishment policy by selecting task-class-specific required fields, listed per class in the \texttt{REQUIRED} map below.

The checker applies a closed-world rule over the required fields:
\[
G(x,z)=
\begin{cases}
\NOTCOMPLETED,\\
\quad\text{if any required field is } \NO;\\
\UNKNOWN,\\
\quad\text{if no required field is } \NO\\
\quad\text{and any required field is } \UNKNOWN;\\
\COMPLETED,\\
\quad\text{if all required fields are } \YES.
\end{cases}
\]
This rule requires evidence about the correct task target and direct support for the task-specific post-condition, rather than support from a proxy cue.

The task class is assigned by a frozen deterministic rule over the task text, not by a learned model or per-record manual labeling. The rule is frozen on Batch~1 before any Batch~2 evaluation and released with record-level class assignments---none manually overridden---so held-out results do not depend on tuning the policy to the validation batch. Each task maps to one of six classes (create/edit, delete, configure, send, query, other); when no class-specific rule applies, the checker falls back to the general \emph{other} obligation set. Given the assigned class, the checker selects the required fields and applies the closed-world rule above:\par
\smallskip

\begingroup
\footnotesize
\begin{verbatim}
REQUIRED = {
  create_edit: [action, target_id, transition,
                committed, final_supports_goal],
  delete:      [action, target_id, transition,
                final_supports_goal],
  configure:   [action, target_id, transition,
                committed, final_supports_goal],
  send:        [action, target_id, committed,
                final_supports_goal],
  query:       [action, target_id,
                final_supports_goal],
  other:       [action, final_supports_goal]
}
\end{verbatim}
\endgroup
\smallskip

\noindent
A missing or negative obligation forces \NOTCOMPLETED{} or \UNKNOWN{}. The checker returns \COMPLETED{} only when every required field is \YES{} with cited trace support, and the model cannot override a failed obligation by asserting a verdict. Each class thereby blocks a characteristic unsupported inference: typed-but-unsaved content for create/edit, an empty list for delete, intermediate state for configure, composed-but-unsent content for send, and a proxy screen for query. The released implementation, schema, and machine-readable checker policy are authoritative; the pseudocode mirrors them.

\subsection{Report--Trace Disposition}

The final layer applies the disposition function \(A\) from Section~\ref{sec:formulation} deterministically, with no language-model adjudication. The overclaim case returns \FLAG, agreement between the two channels returns audit \PASS, and uncertainty in either channel returns \REVIEW.

\subsection{Replayable Evidence--Policy Boundary}

The structured auditor exposes an evidence--policy interface:
\[
\tau \xrightarrow{E} z \xrightarrow{G} T.
\]
Because evidence extraction and policy enforcement are separated, the same cached evidence record can be replayed under an alternative deterministic policy:
\[
T_{G'}(x,\tau)=G'(x,z).
\]
This requires no new model call. An audit operator can therefore ask how decisions would change under a final-state-only policy, a stricter high-precision policy, or a different review-on-uncertainty policy. RQ3 uses this interface to measure precision--load--silent-pass trade-offs under policy mutations.

This replay interface is the main software-engineering difference between the structured auditor and a prompt-level verdict. A \textsc{prompt-blind} judge can be equally report-isolated, but it exposes no per-obligation evidence and cannot be replayed under a new deterministic policy without re-querying the model. The resulting record---evidence fields, cited supports, checker outcomes, and final disposition---can be inspected, regression-tested, and replayed.

\section{Experimental Setup}
\label{sec:setup}

We evaluate the structured auditor as a detector and audit-stage gate for dangerous misreports among audit-target runs. Given an audit-target GUI-agent run \(u=(x,\tau,m)\), the auditor predicts a dangerous misreport when \(m\) conveys unhedged success and the checked trace state is \textsc{not\_completed}. We report results on two record sets throughout the paper. Batch~2 is the held-out validation set, and the combined 465-record set is the complete two-batch analysis over all adjudicated audit-stage records.

\subsection{Corpus and Evaluation Universe}

We collected 1{,}384 in-scope GUI-agent execution attempts under the final harnesses. Our audit applies to report-producing, trace-analyzable runs that coarse harness status cannot decide. The excluded runs are either clean passes, which the harness resolves through its status and report-integrity checks, or non-auditable runs that hit the max-round limit, time out, raise an execution error, or return a missing or invalid report.

Annotation of the remaining runs yields 465 records, 180 of them dangerous misreports under the strict causal-accomplishment label. We refer to these 465 records as the \emph{audit-target set}. All detector metrics in this paper are computed over this set, which is the audit stage rather than the full in-scope stream.

This denominator is the point at which the auditor is invoked, not a post-hoc difficulty filter. Every report-producing, trace-analyzable run left unresolved by coarse harness status enters the audit stage before any auditor prediction is made. Thus, precision and recall measure the component exactly where its decision is used. Full-stream workload is accounted for separately by the audit-stage invocation rate and the gate's \textsc{flag}/\textsc{review} counts, rather than by mixing upstream-resolved runs into detector metrics.

\begin{table}[t]
\centering
\footnotesize
\renewcommand{\arraystretch}{0.95}
\caption{Composition of the 465-record audit-target set. POS = dangerous misreports.}
\label{tab:corpus-composition}
\begin{tabular}{@{}lrrrr@{}}
\toprule
\textbf{Slice} & \textbf{Total} & \textbf{POS} & \textbf{NEG} & \textbf{Med.\ steps} \\
\midrule
\multicolumn{5}{@{}l}{\textit{By source}}\\
AndroidWorld original  & 151 & 78  & 73  & 8  \\
AndroidWorld extension & 276 & 88  & 188 & 11 \\
Themis                 & 38  & 14  & 24  & 17 \\
\midrule
\multicolumn{5}{@{}l}{\textit{By agent}}\\
appagent               & 177 & 90  & 87  & 9  \\
mav3                   & 288 & 90  & 198 & 11 \\
\bottomrule
\end{tabular}
\end{table}

Table~\ref{tab:corpus-composition} shows the source and agent composition of the audit-target set. The set spans AndroidWorld original, AndroidWorld extension, and Themis tasks~\cite{su2021themis} across two agent architectures, AppAgent~\cite{zhang2023appagent} and Mobile-Agent-v3 (mav3)~\cite{ye2025mobileagentv3}, with 90 dangerous-misreport positives from each agent. Themis traces are the longest by median step count, reflecting the more complex report--trace cases retained in the set.

\subsection{Validation Split and Development Protocol}

The evaluation uses two independently maintained data batches. Batch~1 contains 335 records with 130 positives and is used for specification and development. Batch~2 contains 130 records with 50 positives and is held out from prompt, schema, parser, checker, and policy development. Batch~2 contains all Themis records plus additional AndroidWorld records. Since all Themis records are assigned to Batch~2, the auditor was developed on Batch~1 without any exposure to Themis traces. The validation batch therefore includes both held-out AndroidWorld records and a source-level held-out set from a source absent from development.

Batch~2 is a predefined provenance split whose membership is fixed before any model scoring and constrained to preserve group separation and source/action coverage. To make the split auditable, the artifact releases the record-ID lists, a record-level metadata index, and scripts that validate composition, coverage, counts, and group contamination.

The final auditor specification is fixed before scoring. The evidence schema, task-class obligations, deterministic checker, aggregation protocol, and replay policies are not changed after the Batch~2 evaluation.

\subsection{Gold Labels}

Gold labels are assigned on two independently annotated axes. The actual-completion label \(y^*\) is assigned from the task and trace only; annotators do not read the final report when judging completion. The claim-intent label \(c^*\) is assigned from the final report only; annotators do not read the trace when judging what the agent claims. A dangerous misreport is derived as
\[
c^*=\SUCCESS
\quad\wedge\quad
y^*=\NOTCOMPLETED,
\]
not directly assigned as a binary target label.

Two annotators independently labeled each axis after calibration. Neither annotator was involved in developing the auditor's prompts, evidence schema, or audit policies. The protocol uses the causal-accomplishment construct. A task counts as completed only when the trace supports that the agent produced the task-relevant outcome during the run. Boundary rules cover non-existent targets, null observations, pathological repetition, pre-existing artifacts, task-value mismatch, and proxy UI cues, clarified through calibration and disagreement analysis. Before third-party adjudication, inter-annotator agreement was \(\kappa=0.78\) for actual completion, \(\kappa=0.80\) for claim intent, and \(\kappa=0.79\) for the derived dangerous-misreport label. Remaining disagreements were then adjudicated by a third party, yielding the final 465-record gold set. The artifact documents the full protocol, boundary anchors, and disagreement resolution, and includes the annotation/adjudication sheets and manifest.

\subsection{Models, Inference, and Metrics}

We evaluate two hosted model backends, Qwen3.5-35B-A3B and DeepSeek-284B-A13B, denoted qwen and deepseek. Their results are reported separately and never pooled. All model calls use temperature~0. Because hosted APIs can still exhibit residual nondeterminism, multi-run methods are executed five times and aggregated with a 3-of-5 majority vote. If no label reaches majority, the component returns \textsc{unknown} or \textsc{review}; the same protocol applies to the structured auditor and all multi-run baselines, so aggregation is identical across methods. All raw evidence records are retained for auditability.

We report standard detection metrics (precision, recall, F1, and TP/FP/FN counts) together with operating-point metrics (review count, inspection load, blocked recall, silent-pass rate, and inspection yield). Confidence intervals are nonparametric 95\% bootstrap intervals over fixed cached predictions (prediction files and scripts in the artifact). In the automatic-detection view, only \textsc{flag} counts as a positive prediction; \textsc{review} is counted as a non-detection. In the deployment-gating view, both \textsc{flag} and \textsc{review} block automatic pass, so only auto-passed positives remain silent false passes. We define inspection yield as \(\mathrm{Blocked\ POS}/(\textsc{flag}+\textsc{review})\), the fraction of blocked records that are true dangerous misreports.

The auditor is designed for offline use at the audit stage. Auditing one record requires five claim-blind evidence-extraction calls per backend and one shared claim-intent call, matching the call budget of the multi-run baselines. Serialized traces fit within the model context window without truncation: the median trace is approximately 89K characters, or 22K tokens under a four-characters-per-token approximation, and the 90th percentile is approximately 208K characters, or 52K tokens. Outputs are also well below the generation limit: the longest generated output is 1,337 characters, or about 334 tokens, far below the 4,096-token cap. Across the 23,500 model calls in the evaluation, 99.62\% parse cleanly; the remaining 0.38\% are API-empty responses or format violations recovered by a fallback parser, with none attributable to truncation. For the structured extractor itself, parse failures are rare: 1 of 2,325 outputs for qwen and 2 of 2,325 for deepseek.

\subsection{Baselines and Controls}

We compare against simple baselines, claim-aware judges, a prompt-level \textsc{blind} control, and an adapted trajectory judge from prior work. We include three simple baselines---trust-claim / never-flag, flag-all, and claim-gate-only (flag every confident success claim)---to test whether the result can be explained by base rate or claim intent alone. Naive claim-aware judging represents the one-shot pattern that reads the task, trace, and final report together. Matched claim-aware judging uses the same claim gate and five-run aggregation as the structured auditor, but estimates trace completion with a claim-aware judge rather than claim-blind evidence and deterministic checking.

\textsc{blind} is the report-withheld condition used in RQ1. \textsc{prompt-blind} is the report-isolated prompt-level control used in RQ2--RQ3; it withholds the report and emits a prompt-level trace-state verdict but exposes no evidence--policy boundary. The AER-prompt comparator adapts the published Agent-Eval-Refine trajectory-judging prompt~\cite{pan2024autonomous}, preserving its judging role and completion rubric and replacing only the trajectory format with our text-serialized GUI trace.

All LLM-judge baselines receive the same task text and serialized trace as the structured extractor; the only intended information-flow difference is report exposure (claim-aware baselines additionally receive the agent's self-report). All baseline outputs are parsed into the same trace-state label space and evaluated with the same claim gate and five-run aggregation where applicable; full prompts, schemas, and parser rules are in the artifact.

\section{Evaluation}
\label{sec:evaluation}

The evaluation is organized around three research questions. RQ1 tests whether final-report exposure opens an invalid information path into trace-state judging. RQ2 evaluates the fixed structured auditor as an automatic detector and deployment gate, asking whether prompt-level isolation alone is enough or whether structure adds. RQ3 asks what policy replay over cached evidence reveals and where the remaining silent false passes come from.

\subsection{RQ1: Does Report Exposure Bias Trace-State Judging?}
\label{sec:rq1}

RQ1 uses an \textsc{aware}/\textsc{blind} report-exposure intervention in which the same trace-state judge reads the same task and trace. The agent's self-report is appended in \textsc{aware} and withheld in \textsc{blind}; model, trace serialization, scoring rule, and five-run aggregation are held fixed.

Table~\ref{tab:rq1-exposure} shows that report exposure shifts trace-state judgment in the harmful direction. On the combined set, withholding the report improves trace-only dangerous-misreport recall by 43.4 percentage points with qwen and 13.3 percentage points with deepseek; Batch~2 preserves the same direction, with gains of 38.0 and 14.0 points. The flip counts are asymmetric---report exposure mostly suppresses otherwise-correct non-completion judgments. Exact sign tests over harmful versus recovery flips are significant in three of four settings and directionally consistent in the smaller deepseek setting on Batch~2.

\begin{table*}[t]
\centering
\footnotesize
\caption{Report-exposure sensitivity. \textsc{aware} appends the agent's self-report (final report and per-step self-reports), \textsc{blind} withholds it. F1 (R): claim-gated detector view; trace-only \(\Delta R\): recall shift before claim gating; flip (p): two-sided sign test over harmful vs.\ recovery flips.}
\label{tab:rq1-exposure}
\begin{tabular}{@{}llcccccc@{}}
\toprule
\textbf{Scope} & \textbf{Judge} &
\textbf{\textsc{aware} F1 (R)} &
\textbf{\textsc{blind} F1 (R)} &
\textbf{\(\Delta R\)} &
\textbf{Trace-only \(\Delta R\)} &
\textbf{Harmful / recovery} &
\textbf{Flip (\(p\))} \\
\midrule
Combined & qwen &
0.37 (0.23) & 0.62 (0.49) & +0.26 & +0.43 & 79 / 1 & \(<0.001\) \\
Combined & deepseek &
0.59 (0.44) & 0.64 (0.52) & +0.08 & +0.13 & 31 / 7 & \(<0.001\) \\
Batch 2 & qwen &
0.33 (0.20) & 0.58 (0.42) & +0.22 & +0.38 & 19 / 0 & \(<0.001\) \\
Batch 2 & deepseek &
0.59 (0.46) & 0.63 (0.52) & +0.06 & +0.14 & 9 / 2 & 0.065 \\
\bottomrule
\end{tabular}
\end{table*}

\paragraph{Answer to RQ1}
Report exposure creates self-report leakage into trace-state judging, pulling judgments toward \textsc{completed} and suppressing dangerous-misreport recall. This supports the information-flow requirement of Section~\ref{sec:formulation}: the final report must stay out of trace-state estimation. The effect is larger for qwen (+43.4 pp) than for deepseek (+13.3 pp), but both backends lose recall under report exposure, and the held-out Batch~2 set preserves the same direction. The downstream fix does not depend on the larger-effect model---the structured auditor improves dangerous-misreport recall and F1 on both backends (Section~\ref{sec:rq2}).

\subsection{RQ2: Is Prompt-Level Isolation Enough, or Does Structure Add?}
\label{sec:rq2}

RQ1 establishes that the report must be isolated. RQ2 asks what structure adds once \textsc{prompt-blind} already enforces that isolation. Table~\ref{tab:rq2-main} reports the main automatic-detection results. On the combined set, claim-gate-only reaches recall 0.77 but precision only 0.45, because it flags supported and unsupported success claims alike. Matched claim-aware judging and \textsc{prompt-blind} (Section~\ref{sec:setup}) are stronger controls. Consistent with RQ1, matched claim-aware judging trails both isolated methods on recall, so the decisive contrast is the structured auditor versus \textsc{prompt-blind}: both isolate the report and differ only in structure. \textsc{prompt-blind} tests isolation alone, while the structured auditor adds an inspectable, replayable policy boundary. The structured auditor moves the detector toward higher recall and F1---0.75 with qwen and 0.78 with deepseek on the combined set. The corresponding TP/FP/FN counts are 125/28/55 (qwen) and 143/44/37 (deepseek).

\begin{table}[t]
\centering
\footnotesize
\caption{Automatic dangerous-misreport detection (Batch~2 = held-out validation set). Brackets: 95\% bootstrap CIs for structured-auditor F1; TP/FP/FN in text, full intervals in the artifact.}
\label{tab:rq2-main}
\setlength{\tabcolsep}{3.5pt}
\begin{tabular}{@{}lllccc@{}}
\toprule
\textbf{Scope} & \textbf{Model} & \textbf{Method} &
\textbf{P} & \textbf{R} & \textbf{F1} \\
\midrule
Combined & n/a & Claim-gate-only & 0.45 & 0.77 & 0.57 \\
Combined & qwen & Matched claim-aware & 0.84 & 0.42 & 0.56 \\
Combined & qwen & \textsc{prompt-blind} & 0.85 & 0.49 & 0.62 \\
Combined & qwen & Structured auditor & 0.82 & 0.69 & \textbf{0.75} [0.69,0.80] \\
\midrule
Combined & deepseek & Matched claim-aware & 0.82 & 0.52 & 0.63 \\
Combined & deepseek & \textsc{prompt-blind} & 0.82 & 0.52 & 0.64 \\
Combined & deepseek & Structured auditor & 0.76 & 0.79 & \textbf{0.78} [0.73,0.82] \\
\midrule
Batch 2 & qwen & Matched claim-aware & 0.91 & 0.40 & 0.56 \\
Batch 2 & qwen & \textsc{prompt-blind} & 0.91 & 0.42 & 0.58 \\
Batch 2 & qwen & Structured auditor & 0.85 & 0.70 & \textbf{0.77} [0.67,0.86] \\
\midrule
Batch 2 & deepseek & Matched claim-aware & 0.80 & 0.56 & 0.66 \\
Batch 2 & deepseek & \textsc{prompt-blind} & 0.81 & 0.52 & 0.63 \\
Batch 2 & deepseek & Structured auditor & 0.78 & 0.84 & \textbf{0.81} [0.72,0.88] \\
\bottomrule
\end{tabular}
\end{table}

Combined-set bootstrap CIs for the structured auditor are P [0.75,0.87] / R [0.62,0.76] with qwen and P [0.70,0.83] / R [0.74,0.86] with deepseek (F1 CIs in Table~\ref{tab:rq2-main}), indicating a stable result. The recall advantage over \textsc{prompt-blind} comes from paired positive-class contrasts, not from these single-method intervals. On the combined set, the structured auditor detects 40 additional positives with qwen and 49 with deepseek relative to \textsc{prompt-blind}, while losing 3/0 (qwen/deepseek) and adding 13/23 false positives. On Batch~2 it gains 15/16 positives, losing 1/0 and adding 4/6 false positives. Exact tests over discordant positives are significant in all four model--scope settings (all \(p\le0.001\)). Against matched claim-aware judging, the same paired contrast yields 51/54 additional positives on the combined set, losing 2/4 and adding 16/25 false positives, and on Batch~2 gains 15/16 while losing 0/2 and adding 5/6. These discordant-positive tests are likewise significant in all four settings (all \(p\le0.001\)).

Batch~2 validates the same pattern beyond the development batch; it also contains all Themis records, a source absent from development. There the structured auditor reaches F1 0.77 with qwen and 0.81 with deepseek, ahead of matched claim-aware judging and \textsc{prompt-blind} in both recall and F1. \textsc{prompt-blind} confirms that report isolation is already a strong prompt-level control; the structured auditor matches that isolation while improving recall and F1 and exposing the evidence--policy boundary of Section~\ref{sec:design}.

Single-run structured F1 across the five cached temperature-0 runs is $0.73 \pm 0.01$ for qwen and $0.78 \pm 0.01$ for deepseek, close to the aggregated F1 of 0.75 and 0.78. The advantage over the claim-aware controls therefore already holds within a single run, with aggregation mainly stabilizing the result. Against the AER text-only comparator (Section~\ref{sec:setup}), under the same backbone and aggregation, the structured auditor is correct on far more discordant records over the 465-record set (85 vs.\ 24 for qwen, 76 vs.\ 33 for deepseek; \(\Delta\)F1 +0.28/+0.20, McNemar \(p<0.001\)).

To probe dependence on a particular agent architecture, a leave-one-agent-out check stays close to the full-set F1 for both appagent (0.74/0.77) and mav3 (0.76/0.78), each contributing 90 positives (Table~\ref{tab:corpus-composition}); the result does not hinge on a single agent. By source, the AndroidWorld slice (427 records, 166 positives) is likewise stable (F1 0.77/0.79). The Themis slice (38 records, 14 positives) is too small for a stable per-source estimate---a single record shifts F1 by roughly 0.07---so we report its per-source numbers in the artifact rather than drawing a source-specific conclusion.

Automatic detection treats \textsc{flag} as the positive prediction. In audit-stage deployment, \textsc{review} should also block automatic pass, since uncertainty should not be silently accepted. Table~\ref{tab:deployment} therefore reports the deployment-gating view, where both \textsc{flag} and \textsc{review} block automatic pass. On Batch~2, gating blocks 37 of 50 positives with qwen and 45 of 50 with deepseek, at 45\% and 54\% inspection load. On the combined set it blocks 138 of 180 and 159 of 180, at 48\% and 56\% load. These loads are measured over the audit-target set---the hard report--trace cases that coarse harness status cannot resolve. Over the full in-scope stream, clean passes and non-auditable runs are triaged by harness status before this stage, so the gate forwards only the audit-target subset to human review. Against all 1{,}384 in-scope attempts the resulting inspection load is roughly 16\% and 19\% (qwen/deepseek), well below the 48\% and 56\% measured within the audit stage.

\begin{table*}[t]
\centering
\footnotesize
\caption{Deployment gating at the audit stage: \textsc{flag} and \textsc{review} both block automatic pass. Brackets: 95\% bootstrap CIs for the fixed 3-of-5 predictions; load measured within the audit-target set.}
\label{tab:deployment}
\begin{tabular}{@{}llrrrrccc@{}}
\toprule
\textbf{Scope} & \textbf{Model} &
\textbf{Flag} & \textbf{Review} & \textbf{Load} &
\textbf{Blocked POS} & \textbf{Blocked R (95\% CI)} &
\textbf{Silent (95\% CI)} & \textbf{Yield} \\
\midrule
Combined & qwen & 153 & 72 & 48\% & 138/180 & 0.77 [0.70,0.83] & 0.23 [0.18,0.30] & 0.61 \\
Combined & deepseek & 187 & 72 & 56\% & 159/180 & 0.88 [0.83,0.93] & 0.12 [0.07,0.17] & 0.61 \\
Batch 2 & qwen & 41 & 17 & 45\% & 37/50 & 0.74 [0.61,0.86] & 0.26 [0.14,0.39] & 0.64 \\
Batch 2 & deepseek & 54 & 16 & 54\% & 45/50 & 0.90 [0.81,0.98] & 0.10 [0.02,0.19] & 0.64 \\
\bottomrule
\end{tabular}
\end{table*}

\paragraph{Answer to RQ2}
Report isolation improves over claim-aware judging, but it is not enough by itself. The structured auditor recovers additional positives over \textsc{prompt-blind} on both models and both splits, while producing evidence fields and deterministic decisions that can be inspected and replayed. As a gate, it trades moderate inspection load for fewer silent false passes. This trade-off matches the audit-stage cost asymmetry: a missed dangerous misreport passes silently, whereas a false positive only adds review work~\cite{elbaum2001cost,felderer2014taxonomy}.

\subsection{RQ3: What Does Policy Replay Reveal, and Where Does the Auditor Still Fail?}
\label{sec:rq3}

RQ3 asks two diagnostic questions: what policy replay over cached evidence reveals, and where the remaining silent false passes come from.

Because evidence extraction and policy enforcement are separated, cached records replay under alternative deterministic policies without re-querying the model. We replay three mutations over the cached evidence. The \emph{final-state replay} changes the audited construct from causal accomplishment to final-state satisfaction; it lowers F1 to 0.71/0.72 (qwen/deepseek) and raises silent-pass risk by 0.07/0.13, showing that the causal-accomplishment construct matters. The \emph{low-load replay} shifts the deployment preference toward lower inspection load (by 15/28 records) at the cost of higher silent-pass risk (+0.08/0.13), lowering F1 to 0.66/0.70. The \emph{vote-threshold replay} re-aggregates the same five cached runs under stricter flag requirements. Requiring unanimous flags (5-of-5) helps deepseek cleanly: blocked recall rises from 0.88 to 0.93 on the combined set and from 0.90 to 0.96 on Batch~2, and the silent-pass rate falls from 0.12 to 0.07 and from 0.10 to 0.04, at six points more load and at most 0.02 of automatic F1. For qwen the same tightening is a genuine trade-off: the silent-pass rate falls from 0.23 to 0.16 (combined) and 0.26 to 0.14 (Batch~2), but automatic F1 drops from 0.75 to 0.70 and from 0.77 to 0.65, with higher load. These movements come from changing deterministic policy code over fixed evidence, not from prompting a judge again.

Replacing the fixed claim channel with gold claim labels raises F1 from 0.75 to 0.82 for qwen and from 0.78 to 0.84 for deepseek. After deployment gating, most remaining silent positives are trace-side. Of the 42 qwen silent positives, 6 are claim-side misses, 31 are trace-side misses visible to human readers, and 5 are passive-trace limits; the 21 deepseek silent positives split 6, 12, and 3 the same way. Because most are trace-side cases a human reader can identify from the same trace, the trace estimator is the main place to improve.

To gauge how checkable the structured evidence is, Table~\ref{tab:human-bound} compares the trace channel under gold claim intent against human trace-only readers on the audit-target set. Human readers reach F1 0.90--0.95 from the same passive traces, while the trace channel reaches 0.82/0.84, leaving headroom for the trace estimator. Human trace reading is itself below~1.0, so part of the residual is inherent to passive traces rather than a property of the estimator.

\begin{table}[t]
\centering
\footnotesize
\renewcommand{\arraystretch}{0.95}
\caption{Trace channel under gold claim intent versus human trace-only readers, on the audit-target set.}
\label{tab:human-bound}
\begin{tabular}{@{}lccc@{}}
\toprule
\textbf{Trace reader} & \textbf{P} & \textbf{R} & \textbf{F1} \\
\midrule
Human annotator A          & 0.93 & 0.87 & 0.90 \\
Human annotator B          & 0.91 & 0.99 & 0.95 \\
Human (both agree)         & 1.00 & 0.86 & 0.92 \\
\midrule
Structured / gold $c^*$, qwen     & 0.86 & 0.79 & 0.82 \\
Structured / gold $c^*$, deepseek & 0.79 & 0.90 & 0.84 \\
\bottomrule
\end{tabular}
\end{table}

Table~\ref{tab:action-slices} breaks the structured auditor down by task class. The pattern follows the audit obligations: delete and send tasks often expose clearer transition or commitment evidence, whereas create/edit tasks more often require separating durable changes from intermediate typed or visible content. The longer-trace classes do not show inflated recall, so trace length alone does not explain the result.

\begin{table}[t]
\centering
\footnotesize
\renewcommand{\arraystretch}{0.95}
\caption{Structured auditor by task class on the combined set; Blk.\ R treats \textsc{flag}+\textsc{review} as blocking.}
\label{tab:action-slices}
\begin{tabular}{@{}llrrrrr@{}}
\toprule
\textbf{Model} & \textbf{Class} & \textbf{N} & \textbf{POS} & \textbf{R} & \textbf{F1} & \textbf{Blk.\ R} \\
\midrule
qwen & create/edit  & 154 & 48 & 0.58 & 0.67 & 0.63 \\
qwen & delete       & 60  & 38 & 0.84 & 0.87 & 0.95 \\
qwen & configure    & 95  & 38 & 0.66 & 0.70 & 0.68 \\
qwen & query        & 95  & 34 & 0.71 & 0.79 & 0.82 \\
qwen & send         & 32  & 13 & 0.77 & 0.83 & 0.77 \\
qwen & other        & 29  & 9  & 0.67 & 0.63 & 0.89 \\
\midrule
deepseek & create/edit & 154 & 48 & 0.73 & 0.71 & 0.79 \\
deepseek & delete      & 60  & 38 & 0.84 & 0.87 & 0.95 \\
deepseek & configure   & 95  & 38 & 0.87 & 0.79 & 0.90 \\
deepseek & query       & 95  & 34 & 0.74 & 0.79 & 0.85 \\
deepseek & send        & 32  & 13 & 0.85 & 0.85 & 1.00 \\
deepseek & other       & 29  & 9  & 0.78 & 0.64 & 1.00 \\
\bottomrule
\end{tabular}
\end{table}

The 79 harmful flips in Table~\ref{tab:rq1-exposure} (qwen, combined set) are the positives whose correct trace-only non-completion judgment was suppressed by report exposure. On these 79 records, the claim-blind trace channel---which never reads the report---judges \textsc{not\_completed} on 77, and 67 reach an end-to-end \textsc{flag}; the record-level list ships with the artifact. Two cases make this concrete. In a vacuous-delete case, the task asks the agent to delete duplicate recipes, and the report claims success because there are no recipes to process. The claim channel reads success, but the action, transition, and final-state fields all fail, so the auditor flags the run. In an extractor-induced silent false pass, the report claims an alarm was set and the claim channel reads success, yet the extractor emits every required field as \YES{} even though a claim-blind human reader sees non-completion in the same trace. This extractor error, visible to human readers, dominates the residual trace-side misses.

\paragraph{Answer to RQ3}
Policy replay shows that the audited construct and the load/silent-pass trade-off can be adjusted deterministically over fixed evidence, without re-querying the model. The remaining silent false passes are mostly trace-side cases a human reader can identify from the same trace, so the clearest path to further gains is a stronger trace estimator.

\section{Discussion}
\label{sec:discussion}

The central claim of this paper is not that this particular auditor is universally sufficient. Rather, self-reporting agent evaluation needs a clear information-flow rule: the report may define the claim under audit, but it must not enter the trace-state judgment that verifies that claim. The structured auditor is one operational form of this boundary, and structured evidence is what makes the boundary inspectable and enforceable rather than merely prompted.

\subsection{Implications for Self-Reporting Agent Pipelines}

RQ1 shows that exposing a trace-state judge to the agent's own report shifts judgments toward \textsc{completed} and suppresses dangerous-misreport recall. This is harmful precisely because the report comes from the system under audit.

Report isolation should be more than a prompt instruction. \textsc{prompt-blind} shows the value of withholding the report, and the structured auditor adds the evidence--policy interface of Section~\ref{sec:design}: claim-blind evidence fields, deterministic obligation checking, and a consistency table. This makes the auditor inspectable, regression-testable, and replayable (Section~\ref{sec:rq3}) without re-querying a hosted model.

\subsection{Audit-Stage Deployment Boundary}

The auditor implements report isolation as triage for report--trace cases that coarse status checks cannot resolve. In the audit-stage gating view, where \textsc{flag} and \textsc{review} both block automatic pass, it keeps 74--90\% of dangerous misreports from silently passing at 45--56\% inspection load, including on the held-out set. This trade-off suits offline regression, post-hoc audit, and asynchronous review queues where escalation is acceptable.

The deployment boundary follows the same division of labor as the corpus construction. Coarse status checks handle the easy-to-decide outcomes, while the report--trace auditor is invoked only for runs whose completion support requires semantic trace auditing.

\subsection{Remaining Limitations}
 
The auditor operates over passive, post-hoc traces rather than live execution. It can judge whether a success report is supported by the recorded trace, but it does not actively re-execute the task, query the environment, or verify post-conditions outside the available trace. The design also assumes that traces are independently recorded and available for audit; detecting trace manipulation or omissions in the logging substrate is outside this paper's scope~\cite{michael2020forensic,yagemann2021marsara,paccagnella2020kenny,schneier1999secure}.
 
The evaluation is correspondingly centered on the audit stage. Report--trace auditing is defined for runs whose coarse harness status does not already settle the outcome, and our measurements therefore characterize detector behavior at the point where the auditor is invoked. We report audit-stage load, yield, and silent-pass rates for this component, and separately report the full-stream inspection fraction induced by the observed audit-stage invocation rate.
 
Finally, the results are strongest for Android GUI traces under the studied reporting styles and passive trace representation. Residual silent false passes mostly arise when the trace does not provide sufficiently decisive post-condition evidence or when the evidence extractor fails to surface it. This suggests that further gains are likely to come from stronger trace-side evidence extraction, calibrated \textsc{unknown}/\textsc{review} decisions, and deterministic trace features that supplement model-based extraction.

\section{Threats to Validity}
\label{sec:threats}

\subsection{Construct Validity}

A central risk is label circularity: a target label produced by the audit logic under evaluation. Dangerous misreports are therefore derived from two independently annotated axes---actual completion labeled from task and trace without the report, and claim intent labeled from the report without the trace. The target is never assigned by the detector's decision rule.

``Checkable'' means the audit record is mechanically inspectable and deterministically enforced, so field-level errors are visible rather than hidden inside an opaque verdict. The correctness of the emitted evidence fields is bounded indirectly by the human-bound and residual analyses (Section~\ref{sec:rq3}) rather than validated against field-level gold labels.

A second risk is the completion construct itself. We audit causal accomplishment, the strictest reading; relaxing it to final-state satisfaction (the final-state replay, Section~\ref{sec:rq3}) lowers F1 to 0.71/0.72 (qwen/deepseek) and raises silent-pass risk by 0.07/0.13, confirming that the strict construct provides meaningful protection.

\subsection{Internal Validity}

Improvements could in principle come from repeated model calls rather than from the structured design, but all multi-run methods share the same five-run, temperature-0, majority-vote protocol, and single-run stability holds (Section~\ref{sec:rq2}). The separately maintained Batch~2 guards against development-to-validation contamination; its membership is fixed by provenance and coverage constraints rather than model predictions, and record-level split files and validation scripts ship with the artifact. Hosted-model API drift affects full reruns but not the reported metrics, which recompute from cached predictions and evidence records.

\subsection{External Validity}

The corpus is an audit-target set, not a full deployment stream---the report-producing, trace-analyzable runs that coarse harness status cannot decide, a boundary fixed by the harness rather than chosen to inflate measured performance. We report Batch~2 and combined results separately to distinguish validation from complete analysis.
The setting covers Android GUI-agent traces from three sources, two agent architectures, and six task classes. The leave-one-agent-out analysis (Section~\ref{sec:rq2}) shows no dependence on a single agent, though generalization across app ecosystems, agent families, and trace representations remains future work. Load, yield, and silent-pass rates use the audit-target set as denominator (Section~\ref{sec:discussion}). The intended generalization is architectural: keep the report out of trace-state estimation, and compute the audit decision through an inspectable evidence--policy boundary.

\section{Related Work}
\label{sec:related}

This work connects test oracles, GUI-agent evaluation, and LLM-based trajectory judging. The common question across these areas is how to decide whether an execution should be trusted when direct task-specific ground truth is unavailable. The distinctive issue here is a specific leakage mechanism: in self-reporting agent pipelines, the system under audit also emits the natural-language success claim being evaluated.

\paragraph{Oracles, monitors, and cross-channel consistency}
The test-oracle problem asks how to determine execution correctness without an explicit expected output; metamorphic testing, semantic GUI oracles, and runtime verification address related settings with relations, assertions, temporal properties, or domain-specific monitors~\cite{barr2015oracle, segura2016metamorphic, leucker2009runtime, mariani2018augusto, zhang2025assertion}. Our setting is an oracle problem over natural-language GUI tasks and partially observed traces. The post-condition is not a fixed monitor but must be inferred from the task and checked against recorded execution evidence. The closest software-engineering analogue is cross-artifact consistency checking, such as comment--code inconsistency~\cite{tan2007icomment}, ReCoDe's comparison of textual bug reports with screenshots~\cite{yu2023recode}, and screen-level duplicate detection across bug reports~\cite{yan2024janus,yu2024clustering,fang2025prioritization}. We differ in the risk model. The two channels come from an autonomous execution, and an unsupported success report can become a silent false pass, so consistency is not merely a quality signal but an audit gate.

\paragraph{GUI testing, GUI agents, and benchmark verification}
Mobile GUI testing has long explored applications to find faults and generate tests~\cite{mao2016sapienz}, and recent work increasingly uses LLM-driven agents for GUI exploration and task execution~\cite{liu2024gptdroid,hu2024auitestagent,yoon2024intent,zhao2024guitestingarena,yu2024guardian}. Agent benchmarks such as AndroidWorld, SWE-bench, OSWorld, and AndroidLab evaluate task success through task-specific reward functions, executable checkers, or live-environment feedback~\cite{rawles2025androidworld,jimenez2024swebench,xie2024osworld,xu2025androidlab}. GUI agents such as AppAgent, Mobile-Agent-style systems, UI-TARS, and OS-Atlas combine perception, planning, action, and reporting~\cite{zhang2023appagent,wang2024mobileagent,ye2025mobileagentv3,qin2025uitars,wu2025osatlas}. Our target is complementary. After a report-producing run has completed, and when coarse status checks do not settle completion, we audit whether the final report is supported by the passive trace. Safety and trustworthiness benchmarks for GUI agents study harmful or risky behavior~\cite{lee2024mobilesafetybench,levy2024stwebagentbench,shi2025trustworthygui}; report--trace consistency is the completion-side counterpart.

\paragraph{LLM judges, self-verification, and leakage}
LLM-as-judge methods, part of a broader use of LLMs across testing tasks~\cite{wang2024swtesting}, evaluate generated text, plans, and multi-step trajectories, but are known to exhibit position, verbosity, self-preference, and related judgment biases~\cite{zheng2023judging,li2024generationtojudgment,li2025preferenceleakage,panickssery2024selfpreference,sharma2024sycophancy}. Agent-evaluation frameworks such as AER, AgentBoard, Agent-as-a-Judge, AgentRewardBench, and WebJudge-style evaluation use LLM or agent judges for trajectory success~\cite{pan2024autonomous,ma2024agentboard,zhuge2025agentjudge,lu2025agentrewardbench,xue2025illusion}, with the report available or withheld ad hoc. They treat report exposure as an evaluation setting rather than as an information-flow hazard. We are not aware of prior agent-evaluation work that treats report exposure as an information-flow hazard and combines report isolation with a replayable evidence--policy boundary. Our result identifies a specific leakage mechanism in this family. The judged agent's own success narrative can enter the completion estimate and suppress dangerous-misreport recall. Work on hallucination detection, factuality checking, self-consistency, and self-correction asks whether generated text is factual or internally stable~\cite{ji2023hallucination,manakul2023selfcheckgpt,wang2023selfconsistency,huang2024selfcorrect}; here the reference is an execution trace, and the report must be compared against independently estimated trace evidence rather than used as evidence for its own truth.

\section{Conclusion}
\label{sec:conclusion}

This paper studied \emph{dangerous misreports} in self-reporting GUI-agent pipelines: cases where an agent confidently reports success even though the independently recorded trace does not support causal task accomplishment. Our report-exposure intervention shows that one-shot claim-aware judging admits self-report leakage. Exposure to the agent's own report shifts trace-state judgments toward \textsc{completed} and suppresses dangerous-misreport recall. The resulting rule is simple: claim intent should be read from the report alone; completion should be judged from the task and trace alone.

We implement this separation as a report-isolated evidence--policy auditor. Its audit record can be inspected and replayed under alternative policies without new model calls. The auditor improves dangerous-misreport recall and F1 over both claim-aware and \textsc{prompt-blind} controls. Under audit-stage gating, it keeps 74--90\% of dangerous misreports from silently passing at 45--56\% inspection load within the audit-target set, with the same pattern on the held-out Batch~2.

For self-reporting agents, the report should define the claim under audit, not serve as evidence for that claim. The independent evidence should come from the trace or another external source. A system under audit should not provide evidence for its own success.

\section{Data Availability}
Our anonymized artifact includes labels, splits, masked traces, prompts, schemas, checker code, and cached predictions for reproducing all results: \url{https://anonymous.4open.science/r/ICSE27-3C77/README.md}.

\bibliographystyle{IEEEtran}
\bibliography{references}
\end{document}